\documentstyle[aps,prl,twocolumn]{revtex}
\input psfig
\def\picturewidth{2.7in}
\def\dd{\mskip 1.3mu{\rm d}\mskip .7mu}

\font\df=cmssbx10
\def\ga{\gamma}
\def\crit{_{\rm crit}}
\def\tot{_{\rm tot}}
\def\init{_{\rm init}}
\def\eff{_{\rm eff}}
\def\UL{\thinspace dyn}
\def\UT{\thinspace dyn/cm}
\def\UP{\thinspace dyn/cm$^2$}
\begin{document}
\twocolumn[\hsize\textwidth\columnwidth\hsize\csname@twocolumnfalse%
\endcsname
\title{\bf Dynamically-Stabilized Pores in Bilayer Membranes} 
\author{J. David Moroz and Philip Nelson} 
\address{Department of Physics and Astronomy, University of Pennsylvania} 
\address{Philadelphia PA 19104 USA}
\draft
\date{September 1996}

\maketitle
\begin{abstract}
Zhelev and Needham have recently created large, quasi-stable pores in
artificial lipid bilayer vesicles (Biochim.\ Biophys.\ Acta {\bf1147}
(1993) 89).  Initially created by electroporation, the pores remain
open for up to several seconds before quickly snapping shut. This
result is surprising in light of the large line tension for holes in
bilayer membranes and the rapid time scale for closure of large pores.
We show how pores can be dynamically stabilized via a new feedback
mechanism. We also explain quantitatively the observed sudden pore
closure as a tangent bifurcation. Finally we show how Zhelev and
Needham's experiment can be used to measure accurately the pore line
tension, an important material parameter.  For their SOPC/CHOL mixture
we obtain a line tension of $2.6 \times 10^{\rm -6}$\UL.
\\ \\ {\it Running title:} Dynamically-Stabilized Pores 
\\
\end{abstract}
]
\section{Introduction}

Lipid bilayer membranes have remarkable physical properties.  One of
the most important among these properties is a membrane's resistance
to rupture.  In the body, this resistance is critical to the
maintenance of well defined and properly functioning cells.  Indeed,
when a cell needs to undergo a topological change (as it does during
cell division, cell fusion, endocytosis, and exocytosis) it usually
has to make use of specialized machinery which carries out the change
at the cost of chemical energy.  This cost is largely determined by
the material properties of the lipid membranes in question.
\par
We can quantify a membrane's resistance to rupture in terms of a line
tension ($\ga$), the free energy cost per unit length of exposed
edge. Edges are disfavored due to the high cost of either exposing the
hydrophobic lipid chains to water, or creating a highly-curved rolled
edge to hide them. Many authors have devised ingenious indirect
measurements of $\ga$ in various lipid systems
\cite{Taupin1975a,Harbich1979a,Chernomordik1985a}, but direct
measurement has proven difficult.  Among the biologically-relevant
questions which require such measurements is the variation of $\ga$
with lipid shape \cite{Leikin1987a}.
\par
Recently Zhelev and Needham have found a new technique allowing direct
mechanical measurement of the line energy
\cite{Needham1993a,Zhelev1994a}. In this paper we will present a
new analysis of their experimental data.  The experiment revealed some
surprising qualitative phenomena involving pores, which we will
explain. Briefly (see below), they created long-lived quasistable
pores about a micron in radius. After persisting for up to several
seconds, the pores snapped shut in just one video frame. We will
quantitatively explain the longevity of the pores and their sudden
demise, fitting several quite different events with a common value of $\ga$
and two auxiliary parameters.
\par
To see why long-lived pores are surprising, consider the usual energy
of a circular hole in a flat bilayer membrane~\cite{Taupin1975a}.
This energy can be written as a line tension term which is linear in
the pore radius minus a surface tension term which is quadratic.  The
energy thus has the form
\begin{equation}
E(r)=2 \pi r \gamma - \Sigma \pi r^{\rm 2}
\label{naive.energy}
\end{equation}
which has only one stable minimum (at $r=0$).  There is a critical
radius ($r=\gamma / \Sigma$) above which the pore is unstable to
rupture.  To cross this critical point, the system must surmount a
significant energy barrier ($\Delta E = \pi \gamma^{\rm 2}/\Sigma$).
For typical estimates of the line tension ($10^{\rm -6}$\UL)
thermally-driven rupture thus requires a surface tension on the order
of one~\UT, as observed \cite{Evansxx}. For lower tensions, any
transient pore will reclose rapidly, while for larger tensions it will
grow rapidly and lyse the vesicle; in either case, one does not expect
large stable pores to exist.
\par
Nevertheless, Zhelev and Needham found that pores with a radius of
approximately one micron can remain stably open for several
seconds. In these experiments, giant vesicles are aspirated into the
mouth of a micropipette where they are held in place by suction.  A brief
electrical field impulse is applied across the vesicle by a pair of
capacitor plates.  As a result of this impulse, lipid molecules in the
membrane rearrange around a newly formed pore through a process known
as electroporation \cite{Chang1992a}.
\par
As noted above, the appearance of large stable pores is surprising,
and yet Zhelev and Needham documented over a dozen such events. They
proposed that somehow these events managed to sit on the unstable
equilibrium point of Eq.~\ref{naive.energy} for a long time before
suddenly falling off. Inferring $\Sigma$ and $r$ from the data then
let them find the line tension from $\gamma=\Sigma r$. It seems
unlikely, however, that the membrane would remain in unstable
equilibrium for so long. 
\par
In this paper we will find a feedback mechanism that dynamically
stabilizes pores.  The key to the feedback is the relationship between
the outflow of solution through the pore and the velocity of the
vesicle's leading edge as it is aspirated into the pipette.  The
result is a reduced surface tension which is a function of both the
projected length of membrane in the pipette and the pore radius.  This
reduced surface tension yields a new effective energy which exhibits a
thermodynamically stable pore at finite radius.  The pore exists for
some time before suddenly disappearing.  From the critical conditions
leading to the loss of stability, we will be able to produce estimates
of various parameters in the theory; in particular, we will accurately
determine the line tension of the bilayer membrane.

\section{The Experiment}

\begin{figure}
\centerline{\psfig{figure=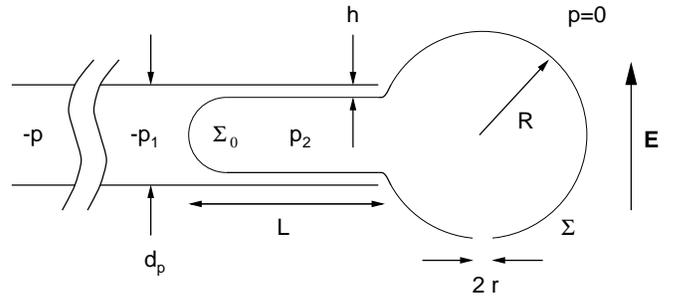,width=\columnwidth}}
\caption{Geometry of the stabilized pore experiment of Needham and Zhelev.}
\label{diagram.notation}
\end{figure}

Figure \ref{diagram.notation}{} defines our notation. In the
experiment lipid bilayer vesicles are prepared from
stearoyloleoylphosphatidylcohline (SOPC) with 50~mol\% cholesterol
(CHOL) at $24^\circ$C. The surrounding solvent is about 0.5M glucose
solution, which effectively prohibits permeation of water through the
membrane by the osmotic clamp effect.  A micropipette is used to
immobilize a chosen vesicle using a suction ($-p$). The suction
pressure is held constant throughout the experiment at a distant
manometer. Initially a small amount of membrane is pulled into the
micropipette, leaving a tense spherical outer bulb of radius $R\init$.
A square-wave electric field pulse is then applied across the vesicle:
typical pulses produce a field on the order of $100\,$kV/m for a
duration of about $150\,\mu$s.  The effect of this field is to
rearrange lipid molecules at one of the vesicle's poles so as to open
a pore in the membrane.  Sometimes no large pore opens; in these
cases the suction is stepped up and a second pulse is applied.  Other
times, the electropore is so large that the whole vesicle gets sucked
rapidly through the micropipette and disappears.  But occasionally the
pore stabilizes and the vesicle moves slowly down the pipette in a
controlled fashion.\footnote{Optical contrast methods make visible the
jet emerging from the pore, and assure us that the pore does close,
rather than just getting pulled into the
micropipette~\cite{Needham1993a}.}
\par
\begin{figure}[t]
\centerline{\psfig{figure=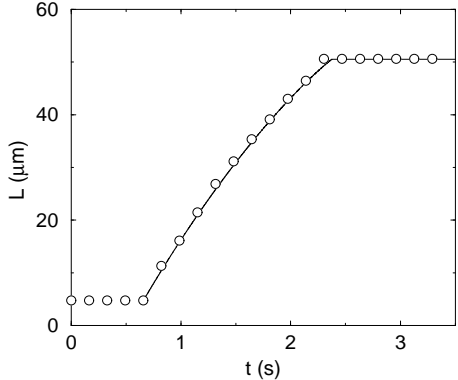,width=\picturewidth}}
\caption{Progress of the leading edge of a
SOPC/CHOL mixed lipid vesicle down the pipette (reproduced from figure
3 of \protect \cite{Needham1993a}).  For this event $R_{\rm init}=17.6 \,\mu$m
and $p=353$\UP.  The initial velocity ($v_{\rm init}$) is
approximately $35\, \mu$m/s.  The solid line depicts the theoretical
$L(t)$ curve for our stabilized pore model.}
\label{time.graph}
\end{figure}

What is measured is then the constant applied suction ($-p$) at the
manometer, the initial bulb size ($R\init$), the micropipette diameter
($d_{\rm p}$), and the location ($L(t)$) of the leading edge of the
membrane as it advances down the micropipette. Figure \ref{time.graph}
shows a typical time course. Other quantities in figure
\ref{diagram.notation}, such as the bulb radius ($R$), the pore radius
($r$), the lubrication layer thickness ($h$), the surface tensions
($\Sigma,\ \Sigma_0$), and the pressures ($-p_1,\ p_2$) are all time
dependent and must be inferred from the directly measured data.

\par
As the vesicle moves into the pipette the suction immediately in front
of it is reduced due to Poiseuille loss along the micropipette.  The
corrected pressure is given by $-p_{\rm 1}$:
\begin{equation}
-p_{\rm 1}=-p + {{32 \eta} \over {d_{\rm p}^{\rm 2}}} (L_{\rm eff}-L) v \ .
\label{poiseuille}
\end{equation}
Here $\eta$ is the viscosity of the sugar solution, $L$ is the
projected length of membrane in the micropipette (figure
\ref{diagram.notation}), and $v\equiv\dot L$ is the velocity of the
vesicle's leading edge.  Eq.~\ref{poiseuille} should really be
regarded as a definition of the effective length ($L_{\rm eff}$) since
the micropipette is not really a perfect cylinder of constant
diameter.  Zhelev and Needham estimated this parameter as $922\, \mu$m
by noting the velocity at which small beads moved down the tube under
similar applied pressures.  Given the significant difference between
this experiment and the one in question, we will treat $L_{\rm eff}$
as an undetermined experimental parameter of fixed
value.\footnote{$L\eff$ also includes any other constant friction, for
example membrane drag around the lip of the micropipette.}{} As we
shall see later, $L_{\rm eff}$ can be determined from the data of
figure \ref{time.graph} using our theoretical approach.
\par
There is a second velocity-dependent friction term that enters into
the equation of motion for the vesicle's leading edge.  This term is
due to shear in the lubricating layer
sandwiched between the membrane and the micropipette wall. This
frictional force creates a
difference between the surface tension ($\Sigma_0$) at the leading edge
and ($\Sigma$) on the exterior bulb:
\begin{equation}
\Sigma=\Sigma_{\rm 0}-{{\eta L} \over {h}}v \ .
\label{friction}
\end{equation}
Here $h$ defines the thickness of the lubrication layer.\footnote{In
reality this parameter will be a function of distance along the
micropipette and of the projected length itself (R.~Bruinsma, preprint
1996).  It turns out however, that the final results are not strongly
dependent on $h$, and so we will treat it as a constant.}  An
experimental determination of this parameter is somewhat difficult;
Zhelev and Needham estimate it to be on the order of $0.2-0.3\, \mu$m.
As with $L_{\rm eff}$, we will determine this difficult to measure
parameter directly from the data in figure~\ref{time.graph}; our value
agrees well with the experimental estimate.

\par
These two sources of friction control the speed of the vesicle front.
As we will show in the following sections, they also determine the
stable pore size and when and if the pore recloses so as to stop the
inhalation process.  As the projected length of membrane inside the
micropipette grows, so does the amount of friction.  We will identify
a feedback mechanism through which the velocity, and hence the surface
tension, becomes a function of the pore size.  This modifies the
effective energy function (Eq.~\ref{naive.energy}) and in doing so,
generates a second stable minimum at a finite pore radius ($r_{\rm
stable}$).  At $L\crit$ the stable minimum disappears and the pore
snaps shut.

\section{Stabilization Mechanism}

The feedback mechanism mentioned above requires the surface tension on
the exterior bulb to be a function of the projected length of membrane
in the pipette and the pore size.  The first step is to write this
tension in terms of the pressure inside the bulb ($p_{\rm 2}$) using
the Laplace formula:
\begin{equation}
p_{\rm 2}={{2 \Sigma} \over {R}} \ .
\label{surface.bulb}
\end{equation}
Here $R$ is the radius of the bulb.  This formula really only applies
at equilibrium; fortunately the membrane's fast relaxation\footnote{A
rough estimate of the relaxation time is given by $\eta_{~\rm lipid}
r^{2}/\gamma \approx 0.01$s; here $\eta_{~\rm lipid} \approx
1$erg~s/cm$^{3}$ is the two-dimensional viscosity of the membrane.
The pore size ($r$) provides a measure of the disrupted area while the
line tension ($\gamma$) gives the restoring force.}  and the fact that
$R$ and $p_{\rm 2}$ change rather slowly imply that the surface is
never far from this ideal.
\par
The radius of the bulb $R$ is already a simple function of $L$.  The
necessary relation is obtained from the constraint of fixed vesicle
area.\footnote{Strictly speaking, the area is not exactly fixed:
thermal fluctuations can always be flattened out by tension to produce
a larger projected membrane area. For the tensions appropriate here,
the area change will not be more than a few percent \cite{Evansxx}.}
We thus have an equation that determines the bulb radius as a function
of projected length ($L$):
\begin{equation}
4 \pi R^{\rm 2} + \pi d_{\rm v} L = 4 \pi R_{\rm init}^{\rm 2}\ .
\label{R.vs.L}
\end{equation}
Here $R_{\rm init}$ is the radius of the bulb at the time the pore is
created while $d_{\rm v} \equiv d_{\rm p} - h$ gives the diameter of
the cylindrical portion of the vesicle inside pipette.
Eq.~\ref{R.vs.L} assumes the bulb region remains spherical and that
the area of the leading edge does not change significantly as $L$
increases.
\par 
The pressure inside the bulb must still be found in terms of $L$.
This is achieved by considering the interface at the leading edge of
the vesicle.  Here we can apply Laplace's formula again:
\begin{equation} 
p_{\rm 1}+p_{\rm 2}={{4 \Sigma_{\rm 0}} \over {d_{\rm v}}}\ .
\label{surface.meniscus}
\end{equation}
Implicit in this equation is the assumption that the pressure is
uniform everywhere inside the vesicle.  There will in fact be a small
pressure gradient near the pipette mouth due to convergent flow.  This
pressure change is on the order of $\eta v/d_{\rm v}$ and turns out to be
negligible.  Eqs.~\ref{surface.bulb} and \ref{surface.meniscus} can
now be used to eliminate $p_{\rm 2}$.  In addition, we can use
Eq.~\ref{friction} to remove $\Sigma_{\rm 0}$ and Eq.~\ref{poiseuille}
to get rid of $p_{\rm 1}$.  This gives us a new form for the surface
tension:
\begin{equation}
\Sigma = {{2R} \over {2R-d_{\rm v}}} \left( {{{p d_{\rm v}} \over {4}} -
\eta v \left( { {{8 (L_{\rm eff}-L) d_{\rm v}} \over {d_{\rm p}^{2}}} + {L \over h}}
\right) } \right)\ .
\label{surface.tension}
\end{equation}
This equation for the surface tension now has explicit dependence on
the projected membrane length (both through $L$ itself and through the
function $R(L)$).  The dependence on the pore size enters only
implicitly through the velocity.
\par
The goal is now to find a suitable expression for the velocity to
substitute into Eq.~\ref{surface.tension}.  Following
\cite{Needham1993a}, this can be achieved by considering the flow
through the pore.  As the vesicle moves into the pipette, its volume
changes: solvent must exit through the open pore.  We
compute the rate of outflow ($Q$) in two different ways and compare to
get $v$.  The first way is to write $Q$ as the derivative of the total
vesicle volume (Figure~\ref{diagram.notation}):
\begin{equation}
Q=-{\dd \over {\dd t}} \left ( {{4 \pi} \over 3} R^{\rm 3} + {{ \pi
d_{\rm v}^{\rm2} L} \over 4} \right) = {{\pi d_{\rm v} v} \over 4}
(2R-d_{\rm v})\ .
\label{flow1}
\end{equation}
Note that we have re-expressed all the time derivatives in terms of
the velocity.  The $R$ time derivative was obtained by differentiating
the area constraint (Eq.~\ref{R.vs.L}):
\begin{equation}
{{\dd R} \over {\dd t}} = - {{v d_{\rm v}} \over {8 R}}\ .
\label{dR.vs.v}
\end{equation}
The other expression for the outflow can be found from the pressure
forcing the flow. Since the flow is at low Reynolds number, $Q$ is
just proportional to the pressure drop across the
pore \cite{Happel1965a}:
\begin{equation}
Q= {{p_{\rm 2} r^{\rm 3}} \over {3 \eta}}\ .
\label{flow2}
\end{equation}
Eqs.~\ref{flow1} and \ref{flow2} can now be combined with
Eq.~\ref{surface.bulb} to yield the desired expression for the velocity of
the leading edge of the vesicle:
\begin{equation}
v = {8 \over {3 \pi \eta d_{\rm v}}} {{\Sigma r^{\rm 3}} \over 
{R(2R-d_{\rm v})}}\ .
\label{velocity}
\end{equation}
This equation can now be substituted back into
Eq.~\ref{surface.tension} to yield the promised form of the surface
tension.  This form has the surface tension as a function of the
projected length (L) and the pore size (r):
\begin{equation}
\Sigma(r,L) = {{\alpha_{\rm 1}} \over {1 + \alpha_{\rm 2} r^{\rm 3}}}\ ,
\label{tension.final1}
\end{equation}
where
\begin{eqnarray}
\alpha_{\rm 1}&\equiv&{{p d_{\rm v}} \over 4} {{2R} \over {2R-d_{\rm v}}} \ , 
\quad \rm {and} \\
\label{tension.final23}
%\label{tension.final2}
%\end{equation}
%\begin{equation}
\alpha_{\rm 2}&\equiv&{{16} \over {3 \pi (2R -{d_{\rm v})^{\rm 2} d_{\rm v}}}} 
\left({ {{8 (L_{\rm eff}-L) d_{\rm v}} \over {d_{\rm p}^{2}}} + { L \over h}} \right) \ . \nonumber
%\label{tension.final3}
%\label{tension.final23}
\end{eqnarray}
These equations define the feedback mechanism through which the
effective surface tension is modified by the friction terms outlined
in the previous section.  The next step is to see how this feedback
mechanism accounts for the experimentally observed behavior --- that
is, to see how it stabilizes pores.
\par
\begin{figure}[b]
\centerline{\psfig{figure=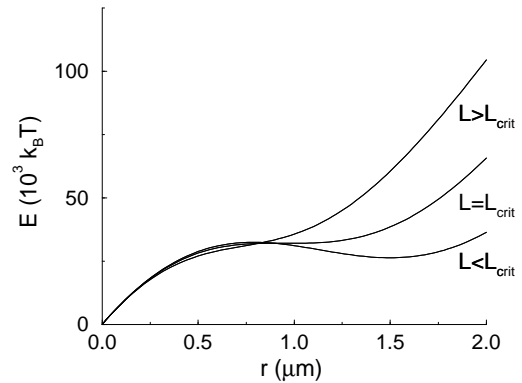,width=\picturewidth}}
\caption{Energy as a function of pore size for supercritical,
critical, and subcritical lengths of aspirated membrane:  $L=0, 45.8, 90
\,\mu $m.  ($L_{crit}=45.8 \,\mu $m.)} 
\label{energy.fig}
\end{figure}

Our main physical hypothesis is now that at each moment the pore
adjusts quickly to minimize an {\it effective energy} similar to
Eq.~\ref{naive.energy}, but with the tension replaced by the varying
quantity just found (Eq.~\ref{tension.final1}). As long as this
effective energy has a nontrivial minimum, the pore size will track
it. This gives the pore size, and hence $v \equiv \dot L$ via
Eq.~\ref{velocity} in terms of $L$.  We can then solve this ordinary
differential equation to obtain the time course $L(t)$ and compare to
the data in Figure~\ref{time.graph}.  This program relies on the
presence of two different time scales: a slow scale for changes of $L$
and $R$, and a much faster time scale on which the pore size $r$
adjusts and the membrane tension equilibrates. We are adiabatically
eliminating the fast variable to obtain a simple dynamics for the slow
one.  Thus our effective pore energy depends on $L$:
\begin{equation}
E\eff(r,L)=2 \pi r \gamma - {{\pi \alpha_{\rm 1} r^{\rm 2}} \over 
{1+\alpha_{\rm 2}  r^{\rm 3}}} \ .
\label{complete.energy}
\end{equation}
This new form of the energy can indeed have two minima: the
trivial one ($r=0$) that appeared before and a new one at a finite
pore size, depending on $L$:
\begin{eqnarray}
\left .{{\dd E\eff} \over {\dd r}} \right |_{r_{\rm stable}} &=&0 \\
\label{first.derivative}
 \rightarrow \gamma &=& 
{{\alpha_{\rm 1}  r_{\rm stable}} \over {2}} 
{{2-\alpha_{\rm 2}  r_{\rm stable}^{\rm 3}} \over 
{(1+\alpha_{\rm 2} r_{\rm stable}^{\rm 3})^{\rm 2}}} \ . \nonumber
%\label{first.derivative}
\end{eqnarray}
As long as $L$ does not exceed some critical value ($L_{\rm crit}$),
there will be a pore size ($r_{\rm stable} (L)$) that satisfies this
equation.  When the projected length reaches this critical value, the
second stable minimum disappears and the pore collapses (figure
\ref{energy.fig}).  At this point the second derivative of the energy 
must also vanish:
\begin{eqnarray}
\left . {{\dd^{\rm 2}E\eff} \over {\dd r^{\rm 2}}} \right |_{r_{\rm
stable,crit}} &=& 0 \\
\label{second.derivative}
\rightarrow 0 &=& 
{{1-7 (\alpha_{\rm 2} r_{\rm stable}^{\rm 3})_{\rm crit} + 
(\alpha_{\rm 2} r_{\rm stable}^{\rm 3})_{\rm crit}^{\rm 2}} \over 
{(1+ (\alpha_{\rm 2} r_{\rm stable}^{\rm 3})_{\rm crit})^{\rm 3}}}\ . \nonumber
%\label{second.derivative}
\end{eqnarray}
The resulting equation is not an equation for the pore size, but
rather for the product $\alpha_{\rm 2} r_{\rm stable}^{\rm 3}$, a pure
number.  The relevant root of equation (\ref{second.derivative})
yields $(\alpha_{\rm 2}r_{\rm stable}^{\rm 3})_{\rm crit} \approx
0.146$.  This number will be useful in the following section where we
will use the critical condition to determine the three undetermined
parameters in the problem: the effective pipette length ($L\eff$), the thickness
of the lubrication layer ($h$), and the line tension ($\ga$) of the membrane.

%The energy function (Eq.~\ref{naive.energy}) can now be rewritten
%in terms of the new surface tension.  The energy thus becomes a
%function of the pore size and the projected length of membrane in the
%pipette.  At this point we need to distinguish two time scales: there is the
%slow time-scale on which the membrane front moves down the pipette and
%the much faster time scale on which the membrane readjusts its
%tension.  Using the large disparity between these two timescales, it
%is possible to eliminate the fast relaxation dynamics of the membrane.
%For each value of $L$, the membrane (and thus the pore size) will
%readjust quickly so as to minimize the energy:
\par
As promised, we have succeeded in reducing the full dynamics of the
problem to only one variable, the projected length ($L$).  This can be
seen by substituting the surface tension formulae
(Eqs.~\ref{tension.final1} and \ref{tension.final23})
%\ref{tension.final2}, and \ref{tension.final3}) 
along with $R(L)$ (Eq.~\ref{R.vs.L}) and the newly determined pore
size ($r_{\rm stable}(L)$) into the velocity (Eq.~\ref{velocity}):
\begin{eqnarray}
\dot L \equiv v(L)&=&{{8} \over {3 \pi \eta}} 
\left ( {{\alpha_{\rm 1}(L) r_{\rm stable}(L)^{\rm 3}} \over 
{1 + \alpha_{\rm 2}(L) r_{\rm stable}(L)^{\rm 3}}}
\right ) \\
\label{velocity.final}
&& \times {{1} \over {R(L) d_{\rm v} (2R(L)-d_{\rm v})}} \ . \nonumber
%\label{velocity.final}
\end{eqnarray}
This equation for $\dot L$ as a function of $L$ can now be integrated
to obtain $L(t)$, then compared directly to the experimental results
(figure \ref{time.graph}).
 %for two column output
% \input{calculation1} %for one column output
\section{Extraction of Parameters}

In the last section we were able to find a stable pore size for each
value of the projected length of membrane in the pipette.  From this
pore size we were then able to determine the velocity,
Eq.~\ref{velocity.final}.  This equation is a first order nonlinear
ordinary differential equation for $L$.  The goal of this section is
to fit the unknown parameters $\ga$, $L\eff$, and $h$ so that the
solution $L(t)$ fits the event shown in figure~\ref{time.graph}. We
will then use these values to explain the other SOPC/CHOL events
documented in \cite{Needham1993a}.
\par
The model that we propose has seven parameters in all (table
\ref{parameter.tab}).  Of these, two are measured directly from
microscope images, namely the initial bulb radius ($R_{\rm init}$) and
the pipette diameter ($d_{\rm p}$).  The pressure ($p$) at the
manometer\footnote{We determined $p$ from \cite{Needham1993a}: the
static tension reported in their table I was substituted into their
Eq.~4 to yield the manometer pressure.} and the viscosity ($\eta$) of
the solvent\footnote{Zhelev and Neeedham used different sugar
solutions inside and outside the vesicle in order to get visual
contrast.  For simplicity, we just used the average of the two
measured viscosities in our calculations.} are also determined
experimentally. \\
\centerline{
\begin{minipage}{\picturewidth}
\begin{table}
\caption{Parameters in the Model}
\begin{tabular}{cc}
Name & Value \\
\hline
$R_{\rm init}$, $p$ & see table \ref{compare.times} \\
$d_{p}$ & $7.5\,\mu $m \\
$\eta$ & $0.014\,$erg~s/cm$^{\rm 3}$ \\
$L_{\rm eff}$, $h$, $\gamma$ & fixed by model \\
\end{tabular}
\label{parameter.tab}
\end{table}
\end{minipage}
}

\par
Of the remaining three parameters,  Zhelev and Needham estimated 
$L_{\rm eff}$ and $h$ via auxiliary experiments, then deduced
$\gamma$.  As discussed in the second section, the effective pipette
length was found using a rather different experiment and so in our
analysis, it will be determined from the data.  Our analysis will also
yield a value for the lubrication thickness ($h$) that will be in good
agreement with the experimental estimate of \cite{Needham1993a}.
Finally we will deduce the last remaining parameter in the model: the
line tension ($\ga$).
\par
\cite{Needham1993a} give the full time course for one event,
which we will use to find the three undetermined parameters listed
above.  This event is reproduced in Figure
\ref{time.graph} and appears as the fourth entry in table
\ref{compare.times} below.  From the experimental $L(t)$
curve we first extract the projected length $L_{\rm
crit}={\df45.8}\,\mu$m at pore closure, the initial velocity $v_{\rm
init}={\df35.5}\,\mu$m/s, and the final velocity $v_{\rm
crit}={\df15.7}\,\mu$m/s. We will now use these numbers to determine
the three unknown parameters: the effective pipette length ($L\eff$),
the lubrication layer thickness ($h$), and the line tension ($\ga$).
\par
The first step is to require that the pore lose its stability at the
observed critical point; the second stable minimum in the energy must
therefore disappear at $L\crit$.  As noted in the last section, the
minimum vanishes when Eqs.~\ref{first.derivative} and
\ref{second.derivative} are satisfied.  To get the pore size just
before closure ($r_{\rm stable,crit}$) we use
Eq.~\ref{velocity.final} at the critical point and require that it
reproduce the observed final velocity.  Recalling
%Eqs.~\ref{tension.final2} and \ref{tension.final3}
Eqs.~\ref{tension.final23} we then obtain: 
\begin{equation}
v_{\rm crit} = { \left ( {{(\alpha_{\rm 2} r_{\rm stable}^{\rm 3})_{\rm crit}} 
\over {1 +(\alpha_{\rm 2} r_{\rm stable}^{\rm 3})_{\rm crit}}} \right ) }
{{p d_{\rm p}^{2}} \over {4 \eta L_{\rm tot}}} \ ,
\label{critical.velocity}
\end{equation}
where we have defined
\begin{equation}
L_{\rm tot} \equiv 8 (L_{\rm eff} - L_{\rm crit}) + {{d_{\rm p}^{2} 
L_{\rm crit}} \over {d_{\rm v} h}} \ . 
\label{Ltot}
\end{equation}
Eq.~\ref{critical.velocity} can be solved for $L_{\rm tot}$.  Then
from Eq.~\ref{tension.final23} and the value of $(\alpha_{\rm 2}
r_{\rm stable}^{3})_{\rm crit}$, one obtains the critical
pore size:
\begin{equation}
r_{\rm stable,crit}= 
{\left( {(\alpha_{\rm 2} r_{\rm stable}^{\rm 3})_{\rm crit} 
{{3 \pi (2R-d_{\rm v})^{\rm 2} d_{\rm p}^2} \over {16 L_{\rm tot}}}
}\right)}^{1/3}
\label{critical.poresize}
\end{equation}
In Eqs.~\ref{tension.final23} and \ref{critical.poresize} the value of
$R$ is to be taken from Eq.~\ref{R.vs.L} with $L=L\crit$.  Using the
critical pore size obtained above, we can now go back to
Eq.~\ref{first.derivative} to obtain a numerical estimate for the
product of the line tension and a yet to be determined multiplier:
\begin{eqnarray}
\gamma {{d_{\rm p}} \over {d_{\rm v}}} &=& {{p d_{\rm p} R} \over {4}}
{\left ( {{ 3 \pi d_{\rm p}^{2} 
(\alpha_{\rm 2} r_{\rm stable}^{\rm 3})_{\rm crit}} 
\over {16 (2 R - d_{\rm p}) L_{\rm tot}}} \right )}^{{1} \over {3}}\nonumber\\
&& \times {{2-(\alpha_{\rm 2} r_{\rm stable}^{\rm 3})_{\rm crit}} \over {(1+(\alpha_{\rm 2} r_{\rm stable}^{\rm 3})_{\rm crit})^{2}}} .
\label{gammaproduct}
\end{eqnarray}
On the right hand side we have made the approximation $(2 R - d_{\rm
v})^{1/3} \approx (2 R - d_{\rm p})^{1/3}$ which is accurate to within
a couple of percent.
%However, it still remains to be shown that
%our model can reproduce the experimental results in detail.  So we
%continue our analysis so as to obtain the values of the other two
%poorly-determined parameters in our model ($L_{\rm eff}$ and $h$).  
\par
The next step is to determine the parameters $L_{\rm eff}$ and
$h$. So far we have found only one combination of these, namely the
quantity $L\tot$ introduced in Eq.~\ref{Ltot}. To separate out the two
contributions to $L\tot$ we now require that the {\it initial}
velocity ($v_{\rm init}$) come out as observed.  From the velocity
formula (Eq.~\ref{velocity}) evaluated at $L=0$, we can find the
effective pipette length in terms of the initial pore size ($r_{\rm
stable,init}$) and the initial velocity:
\begin{eqnarray}
L_{\rm eff}(v_{\rm init},r_{\rm stable,init})&=&{{p d_{\rm p}^{\rm 2}} 
\over { 32 \eta v_{\rm init}}} - {{3 \pi (2R_{\rm init}-d_{\rm v})^{\rm 2} 
d_{\rm p}^{\rm 2}} \over {128 r_{\rm stable,init}^{\rm 3}}} \nonumber \\
&\approx&{{p d_{\rm p}^{\rm 2}} 
\over { 32 \eta v_{\rm init}}} - {{3 \pi (2R_{\rm init}-d_{\rm p})^{\rm 2} 
d_{\rm p}^{\rm 2}} \over {128 r_{\rm stable,init}^{\rm 3}}}. \nonumber \\
&&
\label{Leff}
\end{eqnarray}
This form for $L_{\rm eff}$ can then be substituted back into
Eq.~\ref{first.derivative} to obtain a sextic equation for $r_{\rm
stable,init}$ that can be solved numerically.  The solution so obtained
can be substituted back into Eq.~\ref{Leff} to yield a numerical estimate 
of $L_{\rm eff}$.  Finally the lubrication layer thickness ($h$)
can be recovered using Eq.~\ref{Ltot}.  The approximations considered
in Eqs. \ref{gammaproduct} and \ref{Leff} can now be refined using
this new value of $h$ through a bootstrapping method.
\par
To finish, we recover the line tension from Eq.~\ref{gammaproduct} by 
multiplying through by the factor $(1 - h/d_{\rm p})$.
Thus all three of the required parameters can be determined from the
single aspiration event depicted in Figure~\ref{time.graph}:
$\gamma=2.6 \times 10^{\rm -6}\,$erg/cm, $L_{\rm eff}=313\, \mu $m,
and $h=0.48 \,\mu $m.  This value of $h$ agrees with Zhelev and
Needham's estimate, while our $L\eff$ differs considerably from
theirs.
\par
\centerline{
\begin{minipage}{\picturewidth}
\begin{table}
\caption{Comparison of experimental \protect \cite{Needham1993a} and 
theoretical (this paper) closure times for 
SOPC/CHOL vesicles.} 
\begin{tabular}{ccccc}
$R_{\rm init} (\mu {\rm m})$ & $p$ (\UP) & $t_{\rm exp} ({\rm s})$ & $t_{\rm th}
({\rm s})$ & $Bilayers$ \\
\hline
$18.3$ & $399$ & $-^{*}$ & $2.9$ & $1$ \\
$28.1$ & $310$ & $2.6$ & $3.3$ & $1$ \\
$14.3$ & $331$ & $-^{*}$ & $-^{*}$ & $1$ \\
$17.6^{\ddagger}$ & $353$ & $1.7$ & $1.7$ & $2$ \\
$17.0$ & $324$ & $0.33$ & $0.31$ & $2$ \\
$16.3$ & $337$ & $0.33$ & $0.47$ & $2$ \\
$16.0$ & $355$ & $2.0$ & $1.0$ & $2$ \\
$25.6$ & $168$ & $11.0$ & $11.0$ & $1$ \\
$22.5$ & $240$ & $5.5$ & $5.5$ & $1$ \\
$16.2$ & $320$ & $2.5$ & $4.6$ & $1$ \\
\end{tabular}
\label{compare.times}
\end{table}
$^{*}$ Did not reseal\\
$^{\ddagger}$ Event used to determine $L_{\rm eff}$, $h$, and $\gamma$\\ \\
\end{minipage}
}

With the parameters so determined, it is now possible to determine the
time evolution of the other events in \cite{Needham1993a} from the
initial conditions.  For each of the other nine SOPC/CHOL events
reproduced in table
\ref{compare.times}, we integrated Eq.~\ref{velocity.final} for
the given initial bulb radius ($R_{\rm init}$) and pressure ($p$) to
obtain the critical time at which the pore closed.  In some cases a
double-bilayer with twice the nominal line tension was needed to fit
the data \cite{Needham1993a}.  The table compares the experimental
time to closure found by Zhelev and Needham to the theoretical time
determined using the model.  The same effective pipette length,
lubrication thickness and line tension were used for all ten events.
An effort to optimize the values of the three parameters ($L_{\rm
eff}$, $\gamma$, and $h$) based on a least squares fit to the critical
time data did not produce results significantly different from those
presented above.
\par
Although the parameters were fixed by data from a single event, they
produce reasonable critical times for the entire data set.
The first event in table \ref{compare.times} is clearly an exception
which we have no explanation for.  Perhaps there was a large
fluctuation in the manometer pressure or perhaps the electric field
generated a pore so large that relaxation to the stable pore size was
impossible.  Despite this anomaly we are confident that we have
faithfully determined the line tension of the SOPC/CHOL membrane
making up the vesicles in question.

\section{Conclusion}

In this paper we have explained the existence of the large
dynamically-stabilized pores observed by \cite{Needham1993a}.  By
separating the fast timescale on which the membrane relaxation occurs
from the slow one associated with the motion of the aspirated vesicle
down the pipette, we have been able to establish a modified pore
energy function with a stable minimum in the one micron range.  In
addition, we have described a new mechanism by which this minimum
disappears destabilizing the pore.
\par
The theory that we have developed permits an accurate determination of
an important membrane parameter: the line tension.  From a single
event in Zhelev and Needham's work, we were able to determine this
parameter and two auxiliary parameters.  The values so determined were
then used to reconstruct all ten of the published SOPC/CHOL mixed
lipid events: our theoretical post-prediction for the critical time at
which pore stability is lost agreed well with the experimental result
for all but one event.  The agreement supports the value of our theory
as a method for experimentally determining the line tension of bilayer
membranes.
\\ \\
%%%%%%%%%%%%%%%%%%%%%%%%%%%%%%%%%%%%%%%%%%%%%%
\par\noindent{\bf Acknowledgements:}
We would like to thank R.~Kamien, U.~Seifert, and D.~Zhelev for helpful
discussions, and the referee for suggesting an improvement to our
calculation. This work was supported in part by the  US/Israeli
Binational Foundation grant 94--00190 and NSF grant DMR95--07366.  JDM
was supported in part by an FCAR Graduate Fellowship from 
the government of Quebec.

% \input{conclusion}
% \newpage
% \bibliographystyle{cell}
% \bibliographystyle{authordate1}
% \bibliographystyle{ieeetr}
% \bibliographystyle{jmb}
\bibliographystyle{prsty}
\bibliography{pore}

\end{document}